___________________________________________________________________

# *Single-molecular quantification of flowering control proteins within nuclear condensates in live whole Arabidopsis root.*

Alex L. Payne-Dwyer and Mark C. Leake

Depts. of Physics and Biology, University of York, UK.

## Abstract

Here we describe the coupled standardisation of two complementary fluorescence imaging techniques and apply it to liquid-liquid phase separated condensates formed from an EGFP fluorescent reporter of Flowering Control Locus A (FCA), a protein that associates with chromosomal DNA in plants during epigenetic regulation of the flowering process. First, we use home-built single-molecule Slimfield microscopy to establish a fluorescent protein standard. This sample comprises live yeast cells expressing Mig1 protein, a metabolic regulator which localises to the nucleus under conditions of high glucose, fused to the same type of EGFP label as for the FCA fusion construct. Then we employ commercial confocal AiryScan microscopy to study the same standard. Finally, we demonstrate how to quantify FCA-EGFP nuclear condensates in intact root tips at rapid timescales and apply this calibration. This method is a valuable approach to obtaining single-molecule precise stoichiometry and copy number estimates of protein condensates that are integrated into the chromosome architecture of plants, using confocal instrumentation that lacks *de facto* single-molecule detection sensitivity.

## 1. Introduction

Flowering at the correct time represents a vital evolutionary pressure on a vast majority of commercial crop plants. The timing of each flowering event is largely predicated on exposure to cold winter temperatures, as monitored in *Arabidopsis thaliana* by the regulatory state of Flowering Locus C (FLC) *(1)*. Prior to cold exposure, the FLC gene is repressed by an autonomous pathway, in which 3' processing of antisense transcripts is linked with chromatin silencing markers and transcriptional changes *(2)*. Flowering Control Locus A (FCA) is one of the RNA-processing proteins involved that directly interacts with the polyadenylation machinery at FLC *(3)*. In addition to prion-like domains, FCA contains two RNA recognition motifs which indicate functional self-assembly *(4)*. Indeed, FCA has been discovered to form nuclear condensates in living cells via liquid-liquid phase separation *(5)*.

Phase-separated condensates are strongly evidenced in facilitating gene regulation in a diverse range of cases across living kingdoms *(6–8)*, often as a direct result of their physical properties such as locally increased concentration and buffering of functional molecules *(9)*. The number of active molecules within each condensate is therefore of interest.

An investigator might reasonably then task themselves to detect these nuclear condensates in the living organism and to quantify the local reservoir of functional protein, in this case FCA, within each. However, the nuclear condensates cannot be isolated without perturbation *(10)*, nor are typical commercial microscopes capable of single-molecule quantification at the millisecond timescales on which smaller nuclear bodies diffuse *(5, 11)*. Conversely, single-molecule imaging techniques are limited when applied to multicellular tissues due to scattering and autofluorescence, particularly in plants *(12)*. While liquid-phase separated condensates have been largely investigated using *in vitro* analogues, there remains a keen demand for more accessible experiments *in vivo*.

To overcome these issues, we describe here the coupled standardisation of these two complementary fluorescence imaging techniques. First, we use dedicated single-molecule Slimfield microscopy *(13, 14)* to establish a standard amount of labelled protein in a standard sample; second, we employ confocal AiryScan microscopy to study the same standard; and finally, we interrogate nuclear condensates in whole, wild-type root tips at rapid timescales and apply the calibration. This correspondence between instruments is well-defined since fluorescent protein fusions obtain a consistent brightness under standard imaging conditions *(15)*. The concept is simple: we establish the single-molecule sensitivity of one instrument and lend it to a second microscope which allows superior throughput, ease of use and imaging contrast (Figure 1).

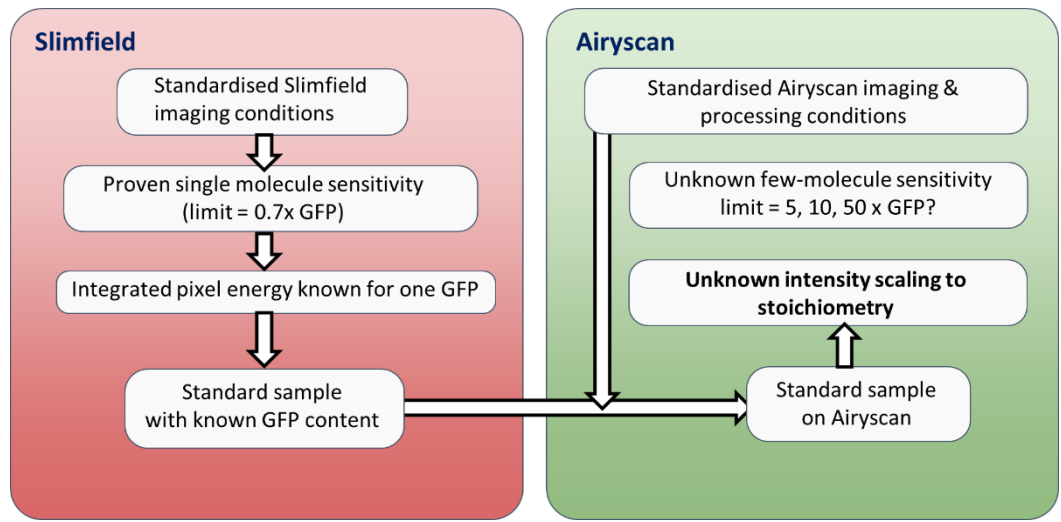

Fig. 1 Flowchart describing the process of lending the single-molecule sensitivity of Slimfield microscopy to the AiryScan confocal microscope suitable for plant root imaging. The process relies on a combination of standardised imaging, a careful choice of standard sample and reproducible analyses.

In our example, the standard sample consists of live yeast cells bearing protein fused to the same EGFP (enhanced green fluorescent protein) label *(16)*. This protein is the metabolic regulator Mig1 which localises to the nucleus under conditions of high glucose *(16–19)*. This approach presents the same chromophore in a similar biochemical environment to the system of interest: FCA-EGFP in the plant nucleus.

Finally, we describe the calibration of the AiryScan root and yeast data using the Slimfield standard, to convert the local fluorescence intensity into the number of labelled proteins per condensate.

## 2. Materials

*Recommended:* A microscope capable of efficient and unequivocal detection of single molecules of EGFP; here we use a custom Slimfield microscope described previously (See Notes 4.4.1 and 4.4.2.) with a high-performance oil immersion objective (Nikon 100x). Again, an 488nm wavelength laser line is required for EGFP excitation.

*Required*: A high-performance confocal microscope (Zeiss LSM880 + AiryScan) with a high magnification immersion objective (Zeiss 63x) and acquisition and processing software (Zeiss, Zen Black). This instrument must be capable of direct imaging of the desired fusion in plant nuclei in live roots. An 488nm wavelength laser line is required for EGFP excitation.

For growing plants:

- *Arabidopsis thaliana* seeds:
    - of a specified accession with stable transgenic fusion: pFCA::FCA-eGFP **(5)** in *Landsberg erecta* **(20)**, and
    - of the control strain lacking the fusion.
- 5% Hypochlorite bleach in uhq $H_2O$
- 1L autoclaved Murashige and Skoog media (4.3 g MS0 powder + 0.5 g MES buffer + 1L uhq $H_2O$, adjusted to pH 5.7 with NaOH)
- 1L autoclaved Murashige and Skoog media as above + 1 %w/w Difco Bacto agar.
- 100mm square petri dishes (plates)
- 300 mL, 2 %w/w agarose in uhq $H_2O$
- 3M Micropore tape
- Aluminium foil
- Sterile razor blades and tweezers
- Laminar flow cabinet
- Cold room or refrigerator at 4°C
- Growth chamber at 22°C (programmable light cycle, humidity and temperature control)
- Racks for holding plates vertically

For growing standard yeast sample:

- *Saccharomyces cerevisiae* (e.g., on frozen stab at -80°C):
    - of a specified background (BY4741) with Mig1-EGFP fusion **(16)**, and
    - of the control strain lacking the fusion.
- Standard petri dishes
- Loose-top vials
- Autoclaved yeast nutrient broth (YNB) liquid media + 0.2 %w/w glucose
- Autoclaved YNB liquid media + 4 %w/w glucose
- Autoclaved yeast extract–peptone–dextrose (YEP) media + 4 %w/w glucose + 1 %w/w agarose)
- Shaking incubator at 30°C
- Plate incubator at 30°C.

For aligning the AiryScan detector and for preparing yeast and plant slides:

- Fluorescent beads matched to AiryScan excitation line:

- TetraSpeck™ Microspheres, 0.1 µm diameter, Invitrogen, 10µL
- Prolong Diamond mountant, 20µL
- Gene Frames 125 µL, Thermo Fisher
- Standard microscope slides, 26 × 76 mm, cleaned
- Oblong coverslips, #1.5 (170 um), 22 × 50 mm, cleaned (see Note 4.3.1)
- An air-fed, benchtop, radio-frequency plasma cleaning unit and vacuum pump (Harrick Plasma), or: 10 mL conc. HCl and approx. 20mL absolute ethanol

For processing:

- A multicore workstation or PC capable of high-performance video processing.
- ImageJ (open-source Fiji distribution)
- PYSTACHIO (Python-based) or ADEMScode (MATLAB-based): particle tracking software suitable for low signal-to-noise input *(20, 21)*.
- CoPro (MATLAB-based) image processing software *(22)*

## 3. Methods

### 3.1 Preparation of live *Arabidopsis* roots for microscopy

The aim of this section is to grow seedlings vertically and when ready, roots are easily mounted on a slide for AiryScan imaging.

1. For each strain, add 20-30 seeds to separate, labelled screw-cap vials (see Note 4.1.1). Add 5% v/v bleach and shake for 5 min. Rinse three times in uhq water.

2. Pour the MS agar medium into six square petri dishes. Seal and leave to set. (See note 4.1.2)

3. When the plates have set, sow the seeds; three plates for each strain. Seal the plates with Micropore tape (See note 4.1.3).

4. Stratify the seeds by storing the plates in darkness (cover with foil) at 4°C for 2-7 days.

5. Grow the plants by setting the plates upright in the growth chamber and leaving them for 5-7 days. (See note 4.1.4).

6. On the day of imaging, slides should be prepared less than 1h before microscopy (see Note 4.1.5). Prepare an agar pad by adhering the Gene Frame to a clean slide. Melt the agarose and mix 500 µL hot agarose with 500 µL of imaging media (for plants, MS liquid media without agar). Fill the well immediately with 500 μL mixture and skim off the excess using a slide to leave a flat agar surface. Wait 5 min for this to set.

7. Select roots that are 1-3 cm long, without mould or damage. Gently pull out each seedling by the leaves using tweezers. Pipette 10μL of liquid MS media onto the agar pad. Cut off the final 8-10 mm of root tip and place this in the centre of the pad. Seal with a large coverslip, taking care to avoid trapping air. Use within 1h (see Note 4.1.5).

### 3.2 Preparation of a standard yeast sample for microscopy

The aim of this section is to grow yeast under controlled conditions including glucose supply, such that the Mig1 protein segregates to the nucleus. The yeast should present as a flat, sparse monolayer suitable for both Slimfield and AiryScan imaging.

1. Thaw each yeast stab and streak under sterile atmosphere onto separate YEP agar plates for each strain. Incubate for 48h at 30°C. (See Note 4.2.1)
2. Seal the plates with Parafilm and store at 4°C (for up to one month).
3. On the day before imaging, set up the clean laminar flow hood. Add 7 ml liquid YNB media at 4% glucose to each of five 15 ml vials for each strain. To the first, add isolated colonies from the corresponding plate.
4. Then vortex to mix, transfer 1 ml into the next vial, and repeat to get successive eightfold dilutions: 1x, 1/8x, 1/64x, 1/1024x, 1/4096x (see Note 4.2.2). Leave these in the incubator overnight (12-18h) (see Note 4.2.3).
5. On the day of imaging, measure the optical densities ($OD_{600}$) and select the culture closest to OD 0.4-0.5. Dilute again 5x into fresh YNB liquid media at 4% glucose and leave for 1h in the 30°C incubator.
6. Construct an agarose pad as above (see section 3.1.6) but using YNB liquid media at 4% glucose instead of MS media. Spot 2μL of culture onto a Gene Frame agarose pad and seal with the coverslip.

### 3.3 Preparation of an AiryScan alignment sample

The aim of this section is to prepare a bright, stable control sample with which the AiryScan excitation laser, pinhole and secondary detector can be mutually aligned.

1. Sonicate the fluorescent beads for 15 min.

2. Place a clean coverslip (see Note 4.3.1) on a clean benchtop and pipette 5 µL undiluted bead stock in the centre. Spread the drop out with the pipette tip and leave to dry for 5 min.
3. Place a clean slide on the benchtop and pipette 10 µL of Prolong Diamond onto the centre of the slide.
4. Add the coverslip with the bead-coated side facing downwards. Leave with the coverslip facing down in the dark for 16h (overnight) to allow the mountant to cure with the beads in place. Store at 4°C (see Note 4.3.2).

**3.4 Optional: Slimfield imaging of the standard sample**

This optional section aims to generate Slimfield photobleaching sequences (see Note 4.4.1) of yeast nuclei in a high glucose condition, with for the purposes of evaluating i) the initial intensity of each yeast nucleus, ii) the characteristic brightness of a single EGFP molecule *in vivo*., and thus iii) the number of molecules in a yeast nucleus. Since literature values of the number of Mig1-EGFP per cell are available (see Note 4.6.3), this section is not strictly required for a basic AiryScan calibration but is recommended for improved accuracy.

1. Align the Slimfield microscope *(20)* (see Notes 4.4.1 - 4.4.3.) with 488nm excitation using the fluorescent bead sample.
2. Mount the Mig1-EGFP yeast slide. Find the cells in brightfield and focus on the cell midbody.
3. Record a brightfield image for later segmentation and inspection.
4. Set the fluorescence acquisition parameters (see Note 4.4.4) such that i) the sample is not initially saturating the camera (Figure 2A, top centre), and ii) when the sample is undergoing photobleaching and almost spent, individual EGFP molecules can be seen blinking across individual frames. Images should be saved as 16-bit grayscale with camera metadata in OME TIFF format.
5. Set the number of frames to acquire sufficiently large to capture this photoblinking regime.
6. Record the standardised excitation and emission settings, including laser power and filter types.
7. Change to brightfield and find a new, pristine yeast cell.
8. Acquire in fluorescence mode with the same exposure and power settings, ensuring that the sample is not exposed to the laser until the first frame of the recording takes place. (See Note 4.4.5).
9. Repeat with the same settings to obtain >50 cells from each of at least 3 independent cultures.
10. Repeat for the unlabelled strain.

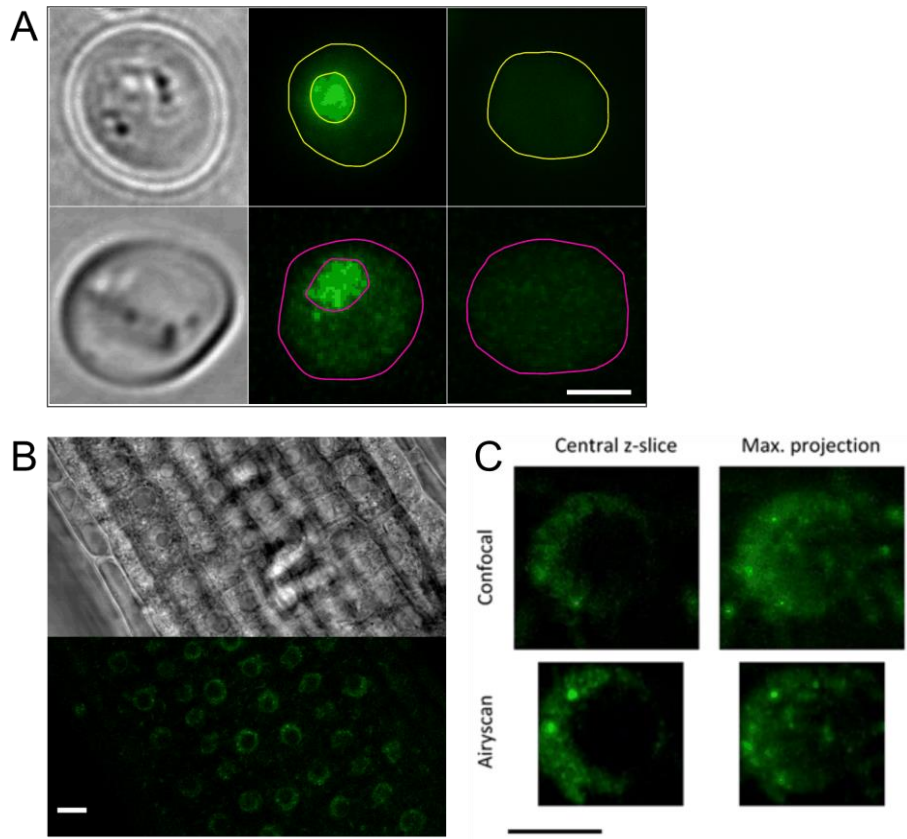

Fig. 2 Representative acquisitions for the Mig1-EGFP yeast culture and FCA-EGFP root. (A) Yeast cells in high glucose condition expressing Mig1-GFP that localises to the nucleus, with segmented nuclear and cellular boundaries where visible; (first row) Slimfield microscopy and (second row) LSM880 confocal after AiryScan processing; (left) transmitted light, (centre) Mig1-EGFP fusion strain, (right) parental control showing autofluorescence. Scale bar 2 µm. (B-C) LSM880 confocal images of an FCA-EGFP labelled Arabidopsis root; B) in zoomed-out overview: (top) transmitted light and (bottom) fluorescence channels; Scale bar 10 µm. C) A single epidermal nucleus, in single slice and maximum intensity projection along z, before and after AiryScan processing. These frames show spots corresponding to biomolecular condensates of FCA-EGFP. The condensates reside in the nucleoplasm surrounding the nucleolus (central dark area). Scale bar 5 µm.

### 3.5 AiryScan imaging of the standard and root samples

This section details the process of aligning the AiryScan instrument (see Note 4.5.1), followed by imaging roots and then the standard yeast sample.

1. Mount the alignment bead sample with the 63x oil immersion objective. Set the AiryScan 'RS' mode and turn on the laser at 1% power. On the Maintenance tab, select 'Adjust in Continuous mode' to allow the pinhole to move in response to the AiryScan detector. Refocus onto the beads in 'Continuous' mode until the signal is centred on the AiryScan detector. Select 'Store position' and deselect 'Adjust in Continuous mode'. The AiryScan is now ready to use for a dim sample.

2. Mount the FCA-EGFP root sample. Use 'Locate' mode to centre the root tip under brightfield illumination. Change to Acquisition tab and use 'overview'-style settings at low zoom to ensure the nuclei are in focus (Figure 2B).
3. Save an acquisition so that these overview settings can be reproduced with the 'Reuse' function in Zen Black. Images should be saved as 16-bit grayscale with metadata in .CZI format.
4. Choose the cell type of interest. These must occur at a consistent layer (epidermal or cortex) and distance from the root tip (meristem, dividing zone or elongation zone). FCA condensates are typically brighter/higher contrast in cortex nuclei in the dividing region.
5. Explore the AiryScan mode settings to determine a suitable standard imaging configuration (see Note 4.5.2 and 4.5.3) which can capture FCA condensates of all sizes and intensities.
6. Fix and record these settings by saving the acquisitions so that they can be recovered by means of the 'Reuse' function.
7. Acquire and save z-stacks of the root nuclei of the desired type using the standard settings. Repeat to obtain >50 cells from 3 independent cultures.
8. Repeat for the unlabelled control strain.
9. Mount the yeast sample. Acquire and save z-stacks of the standard yeast cells using the standard settings (Figure 2A, second row). Repeat to obtain >50 cells from 3 independent cultures.
10. Repeat for the unlabelled control strain.
11. Navigate to the Processing tab. Perform batch-wise 'AiryScan Processing' with a suitable choice of processing 'Strength' to generate super-resolved volumes without artifacts (see Note 4.5.4)

**3.6 Single-molecule brightness using literature estimates**

The purpose of this section is to analyse the newly acquired images to extract the characteristic brightness of a single molecule *(20)* in the Airyscan data, here termed *Isingle(A)*.

1. Collate the AiryScan processing output and, if available, Slimfield image sequences on the workstation.
2. Load an AiryScan z-stack of the standard yeast data into ImageJ using the 'BioFormats' plugin.
3. Perform a maximum intensity projection to collapse all the z-slices (see Note 4.6.1). Save as TIFF.

4. Segment the yeast nuclei to generate binary pixel masks, for example using local thresholding in ImageJ, then iterating to optimise the heuristics (see Note 4.6.2). CoPro software *(22)* requires input as a '_segmentation.mat' MATLAB array file. To generate this, use ImageJ to save as binary TIFF, then use MATLAB to convert this to a binary array and save as the .mat file.
5. Measure the total fluorescence intensity of the nucleus in pixel gray value units using CoPro software with the dummy input Isingle = 1. Use a dark pixel value (typically bias of 100) as the 'background', and the segmentation masks. Use the default option for a simulated point-spread function.
6. The output of CoPro is a .mat file detailing the brightness of each yeast nucleus in terms of counts. This gives the raw integrated count (RIC) for each nucleus. It also gives the size of the segmented area.

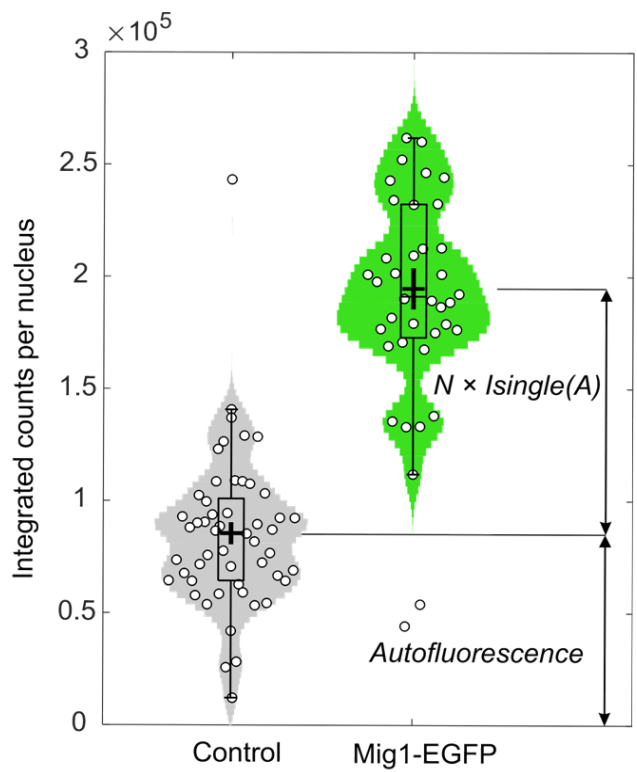

Fig. 3. Violin plots of fluorescent integrated counts (area corrected) of yeast nuclei under Airyscan microscopy as obtained from CoPro with dummy input of *Isingle* = 1 for control and Mig1-EGFP (n=44, 50 nuclei respectively). The individual nuclei are shown as circles, with boxplot overlaid. The difference between population mean counts (black crosses) is proportional to both the mean number of Mig1-EGFP molecules in each nucleus, *N*, and to the characteristic *Isingle(A)*, the number of counts associated with each EGFP in every AiryScan standard acquisition.

7. Repeat for all the other labelled AiryScan yeast sequences; also repeat for the unlabelled parent strain z-stack, where the RIC represents only the total autofluorescence intensity (Figure 3).

8. For each of the two yeast strains, take the mean of the AiryScan RIC over all the corresponding nuclei:

$$\overline{RIC}(labelled) = \frac{1}{n}\sum_{i=1}^{n} RIC\ (labelled, \text{nucleus } i)$$

9. Find the mean copy number of Mig1-EGFP molecules per nucleus, *N*. This can be verified independently using CoPro on the yeast Slimfield sequences if available (see section 3.7 below), though estimates are already known for Mig1-EGFP from published work (see Notes 4.6.3 and 4.6.4).

10. Take the difference of the two average *RIC* values, with the control weighted by the mean cell areas in each strain determined by CoPro. Then divide by *N* to estimate the *Isingle* value for AiryScan:

$$Isingle(A)\ =\ \frac{\overline{RIC}(labelled) - \overline{RIC}(control) \times \frac{Area(labelled)}{Area(control)}}{N}$$

The result should be representative of any EGFP molecule under the same AiryScan imaging conditions, including condensed FCA-EGFP molecules in the plant nuclei.

### 3.7 Optional: verification of single-molecule calibration using Slimfield imaging

The core use of Slimfield microscopy is to estimate the single molecule brightness and the absolute number of localised molecules directly from image sequences. This section is concerned with analysing the Slimfield imaging of the yeast nuclei to refine the *Isingle* calibration above.

1. Use PySTACHIO *(23)* or ADEMScode *(21)* software to track the Slimfield images of yeast with Mig1-EGFP.

2. The integrated pixel value brightness of one EGFP molecule in Slimfield data, *Isingle(S)*, can be estimated quickly by calculating the modal spot brightness after 1/3 of the photobleaching has occurred. This is an inbuilt feature of PySTACHIO and is governed by the 'InVitroIsingle' script in ADEMScode. Alternatively, identification of consistent stepwise photobleaching is a more robust method of finding *Isingle(S)* for Slimfield *(24)*.

3. Segment individual nuclei in the Slimfield data using the inbuilt tools in PySTACHIO or ADEMScode for yeast nuclei. Generate masks for each field of view that will be included in the next step, CoPro.

4. Run CoPro in MATLAB on the Slimfield data, with input including the *Isingle(S)*, a dark pixel value (typically bias of 100) as the 'background', and the segmentation masks. Use the default option for a simulated point-spread function.

5. The output of CoPro is a .mat file detailing the brightness of each yeast nucleus in terms of the number of equivalent EGFP molecules.

6. Repeat steps 3-5 for the parental control strain.

7. Take the average and SEM of the copy numbers in the EGFP-labelled and control cases. Take the difference of these two numbers. This result is *N*, the number of actual molecules of Mig1-EGFP per nucleus. This empirical result can be used in the calculation in section 3.6 instead of the literature value, to refine *Isingle(A)*, the brightness counts corresponding to a single EGFP in the AiryScan data.

8. Ensure to propagate the SEM uncertainty on each value (by addition in quadrature) from the CoPro output, through to the estimation of the number of molecules per nucleus, *N*, and finally *Isingle(A)*, based on the number of nuclei, *n*, included in the calculation. For $n = 50$, the relative error in *Isingle(A)* is expected to fall around 10%.

**3.8 Applying the calibration to analysis of Airyscan plant root images**

In this section the calibration is applied to the AiryScan data for FCA-EGFP bearing plant nuclei. The trajectory analysis divides the integrated counts of each trajectory's initial timepoint by the *Isingle(A)* brightness; this refinement yields stoichiometry not simply per cell or per nucleus, but of FCA-EGFP molecules per condensate focus.

1. Track the roots using PySTACHIO or ADEMScode software to detect foci and link them into trajectories.

2. Determine the cumulative uncertainty in the stoichiometry rescaling by adding in quadrature: i) the fractional error in Isingle(A) to ii) the baseline Slimfield sensitivity of 0.7 EGFP molecules (see Note 4.6.3).

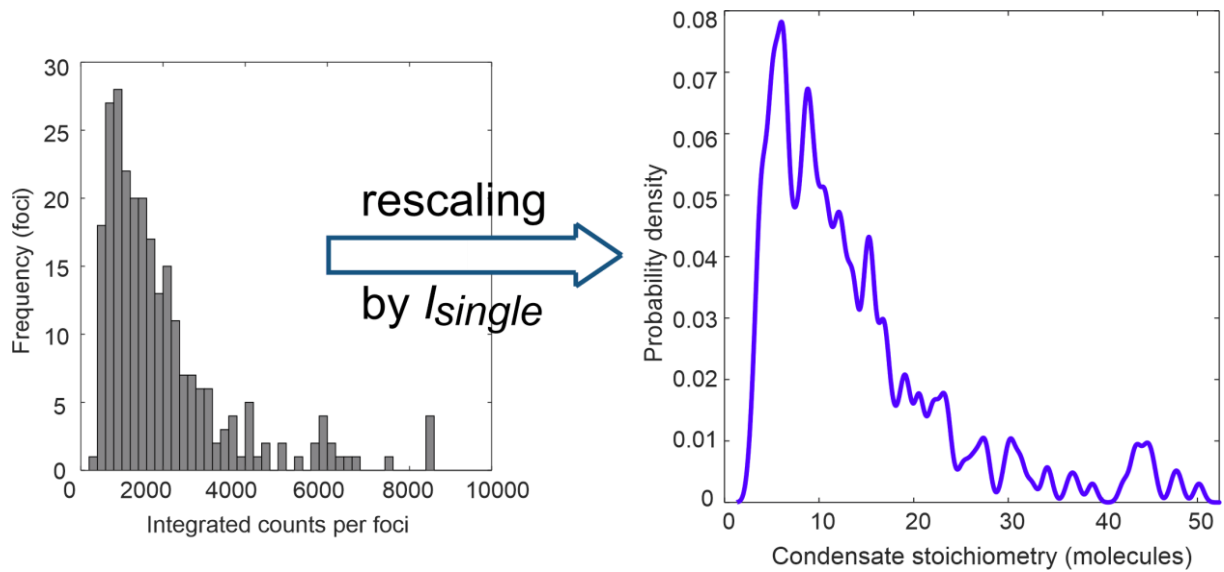

Fig. 4. Stoichiometry of FCA-EGFP assemblies within analysed AiryScan volume stacks in live root tips ($n$ = 2 replicates, 22 nuclei, 163 foci). Scaled from integrated intensity counts by $Isingle(A)$ = 134 ± 15 PGV. Kernel density estimate with kernel of 0.71 molecules.

3. Prepare the trajectory analysis script ('TrackAnalyser' in ADEMScode). Set the input *Isingle* to the to the identified value of *Isingle(A)*, and the kernel density width, *params.stoichKDW,* to the total rescaling uncertainty.
4. The stoichiometry distribution will be displayed as a kernel density estimate (see Figure 4) and saved to the output.mat file.
5. The *Isingle(A)* value can be applied to other datasets for which the settings are maintained – that is to say, the number of counts per EGFP per frame remains consistent with the standard yeast sample (see Note 4.6.5 and 4.6.6).

### 4. Notes

**4.1 Preparation of live roots for investigation**

4.1.1: Dry seeds are very susceptible to electrostatic dispersal, so avoid pouring/handling them and use a long metal spatula instead.

4.1.2: If roots are stressed, such as by dehydration, they may present aberrant phenotypes including high autofluorescence. Ensure the depth of the agar on the plates is at least 8mm (around 70 mL) and that the plates are closed when the agar is cooling.

4.1.3: It is helpful to draw a horizontal line across the top half of the underside of the plate as a guide for sewing. A pipettor is too aggressive; use a 1000 µL pipette tip and a gloved finger to provide fine control, so the wet seeds are dispensed individually. The tape retains water while allowing gas exchange; do not use impermeable tape.

4.1.4: The *Arabidopsis* growth chamber should be set to a fixed temperature, humidity, light intensity, and light duration. Here, 22°C and 50% relative humidity with a 16h day / 8h night cycle is recommended. Vernalisation in a cold chamber may be added after the growth period but is not required for this calibration protocol.

4.1.5: The cut roots are susceptible to osmotic shock, so MS media is preferred over uhq water for imaging. If prepared correctly each root will last for about 1h before degrading; the signature that must be avoided is if the nuclei begin to disperse and debris accumulates inside the cells. Once this occurs discard the roots and cut another set for a new slide. This protocol is intended for short-term endpoint imaging. For timelapse imaging, whole seedlings must be mounted and other influences such as root growth must be accommodated **(25, 26)**

**4.2 Preparation of the standard yeast sample**

4.2.1: Yeast inocula are susceptible to contamination by fast-growing bacteria, so all culturing is best performed in sterile atmosphere.

4.2.2: The serial dilutions must match the incubation time. Based on an optimal yeast doubling time of about 1.5h, then an overnight gap of 12-18 hours is covered by dilutions around $2^{(-12/1.5)} = 1/256$ and $2^{(-18/1.5)} = 1/4096$. Additional variability arises since the agar-grown cultures don't homogenise easily under vortexing and the amount inoculated from the plate into each vial is not easily controlled. Multiple dilutions are necessary to provide contingency.

4.2.3: Pull the lids up after setting the vials in the shaking incubator to make sure they are not airtight, otherwise the culture will become anaerobic.

**4.3 Preparation of an AiryScan alignment sample**

4.3.1: Coverslips must be free of fluorescent contaminants, or the yeast sample will suffer from extraneous background. To clean slides and coverslips in tandem, either use a low-pressure air plasma cleaner (Harrick) for 10 min, or clean the coverslips for 30 min in 1% hydrochloric acid (7 mL conc. HCl in 250 mL uhq $H_2O$ on a shaker). Rinse thoroughly in uhq $H_2O$. Incubate in absolute ethanol on a shaker to 30 min. Leave glass to dry. Store the slips for up to a month in a sealed container such as a Falcon tube or slide box.

4.3.2: Once the bead sample has cured, store it in the fridge for up to one month. Do not freeze it. Vectashield or glycerol can be used as alternative mountants, but the beads will dislodge more easily.

**4.4 Slimfield imaging**

4.4.1: Details of our bespoke Slimfield setup comprising off-the-shelf optics and optomechanics are published extensively elsewhere *(27–35)*. Here, only single colour imaging in GFP channel is required, though the major determinants of performance are i) the correct matching of emission filters to the fluorophore and ii) the quality of the scientific camera (here a Photometrics Prime 95B running at the highest gain setting). Suitable commercial alternatives for microbiological standard samples include TIRF-capable microscopes such as the Oxford NanoImager (ONI).

4.4.2: Slimfield reveals the quantitative characteristics of individual tracked bodies in greater detail than simply the number of foci and their spatial (co)distribution within the cell. Information on the diffusive mobility of bodies and that of their internalised molecules can be obtained, as well as the estimates of stoichiometry and protein copy that are discussed here. Methods to explore the mobility of loci in Arabidopsis using complementary techniques are described elsewhere *(36)*.

4.4.3: While the excitation beam must be collimated (as in a TIRF microscope), the yeast nuclei must be imaged at normal beam incidence since they do not rest on the coverslip. The excitation beam must be wide enough that the illuminated area containing at least one whole nucleus is appreciably uniform. The emission output of one molecule is then approximately the same, regardless of its location in the nucleus.

4.4.4: For all fluorescence-based intensity measurements, there is a trade-off between trajectory length and frame rate. If the laser power is too high, the sample will photobleach very quickly and the tracks of individual molecules will be short. If the power is too low, then the exposure time must be extended (frame rate reduced) to retain single-molecule detection in each frame. For our implementation, the region of interest is 190 × 190 pixels = 10 × 10 microns at 50 nm/pixel; the typical exposure time is 5 ms (189 frames per second); the laser beam is 14 microns wide (Gaussian $1/e^2$ diameter) and the peak intensity is 5 kW/cm$^2$.

4.4.5: The sample can be protected from premature photobleaching either by digital control of the laser or by control of the excitation shutter. The signal will therefore not be photobleached prior to the acquisition and all molecules of Mig1-EGFP will be visible in the first few frames.

### 4.5 AiryScan imaging

4.5.1: Zeiss AiryScan instruments *(37)* use a second photodetector to measure the real confocal point-spread function during the acquisition of each confocal image. This information is then used in postprocessing steps to undo the blur from optical diffraction. A carefully chosen deconvolution step will refine the spatial resolution as low as 150nm.

4.5.2: Protocols are available for generating confocal z-stacks of *Arabidopsis* nuclei *(36)*. However, the requirements for fixing the settings are much stricter here. Once the settings are standardised, they cannot be adapted later to suit unseen samples. Do not change any settings or optics on either the excitation or emission paths, just the slide and the stage and focal positions. These filters, power and exposures must remain fixed, or the subsequent calibration will become invalid; the protocol here would need to be repeated. There is a clear benefit of the 'Reuse' function on a well-maintained microscope with multiple users.

4.5.3: The laser power and pixel dwell time must be adjusted to give best contrast for smaller, diffraction limited FCA bodies (Figure 2C). The region of interest, z-slices and z-spacing must be set so that an entire nucleus is captured to avoid undercounting. A suitable starting point for FCA-EGFP is recommended as follows: AiryScan 'RS' acquisition mode; confocal region of interest: 9 x 9 microns = 100 x 100 pixels at pixel size 90 nm. This reduces to 7.6 x 7.6 microns per slice after AiryScan processing. Laser power: 4% of source max ~ 0.8 mW. Exposure: 56 ms/slice × 15 z-slices at 0.6 microns per step = 0.8 s per volume 9 microns deep. The optimal pixel size is less than the diffraction limit since mild spatial oversampling is preferred for super-resolved tracking of foci. The photobleaching decay time of EGFP under these conditions is about 1 min of exposure ~ 1000 slices, or ~50 cycles.

4.5.4: The Wiener deconvolution used in the AiryScan processing is a nonlinear high-pass filter, so is vulnerable to introducing severe spot-like artifacts at low input SNR. For dim samples, the filter strength needs to be kept low to avoid extra false positives when tracking. Filter strengths of 5.0 are recommended for z-stacks and just 2.3 for still images, instead of the 'Auto' Zen Black settings.

### 4.6 Single-molecule calibration, verification and implementation

4.6.1: The maximum intensity projection is only necessary to collapse the XYZT sequences to representative XYT images, so that all (local maxima) can be tracked, since the tracking software is designed for 2D image sequences over time. Segmentation is also much simpler in 2D. However, this step necessarily makes some assumptions about the sparsity of the fluorophore distribution.

There is a limitation that if the number density of condensates is very high then some local maxima will be lost in the projected and not detected, or their assignment to tracks made more ambiguous. Since the maximal intensity projection also skews the pixel values higher as a function of the number of slices, the number of slices should also be kept constant for the calibration to hold. Alternatively, one could track all the individual frames across z and t, though these would be noisy, computationally challenging and risk overcounting the molecules, were any to drift in z position during the scan.

4.6.2: CoPro *(22)* uses a traditional image processing pipeline, designed for supervised segmentation in the ADEMScode suite that is dependent on manual input and/or optimising heuristics. These heuristics include the number and order of morphological operations, such as closing and eroding, which must be determined by trial and error. ImageJ is also capable of performing segmentation in this way. A benefit of ImageJ and MATLAB is that the interfaces allow intuitive visual assessment of each iterative step or result. However, there is a wealth of alternative, less heavily supervised methods for segmenting 2D images of yeast cells, such as convolutional neural networks which could scale more readily to larger image sets *(27)*.

4.6.3: The standard Slimfield estimate is 806 ± 353 molecules (mean ± SD.) of Mig1-EGFP per cell at high glucose condition, of which 226 ± 155 reside in the nucleus *(16)*. The resulting *Isingle(A)* value is 134 ± 13 counts/molecule EGFP (n=50 cells, 3 replicates). Our abbreviated calibration method relies on this literature estimate and thus assumes the mean total copy of Mig1 protein is consistent and reproducible for each population of yeast cells *(18)*. Ideally one would perform the AiryScan and Slimfield standard acquisitions on the day with the same standard slide(s), then repeat on two more days to generate three biological replicates. However, the variance in the original Slimfield measurement of the stepwise photobleaching of single molecules (0.7 per EGFP) tends to dominate over the uncertainty in the AiryScan measurements.

4.6.4: We do not make a correction for any maturation effect for EGFP, but a separate protocol is available to characterise this if appropriate *(20)* Under the assumptions of fast maturation, robustness to environmental conditions and simple three-state fluorescence behaviour, this protocol may not be limited to EGFP, but may be useful for exploring calibration of other constitutively fluorescent proteins.

4.6.5: The method assumes that the brightness of the fusion protein is indistinguishably similar in the two samples. For deeper imaging within the plant, the scattering loss will increase. Since the yeast are imaged at the surface, they necessarily cannot reflect the reduction in effective molecular brightness as a function of depth. This means the calibration strictly provides a lower bound estimate on the stoichiometry in the plant nuclei, which becomes less accurate with depth.

4.6.6: Applying the calibration provides for single molecule resolution between peaks in the stoichiometry but cannot change the fundamental sensitivity cut-off (noise floor) of the AiryScan instrument. As such, individual, isolated molecules are likely to remain undetected by the AiryScan instrument. This is visible as a steep shoulder at the lower end of the AiryScan-derived stoichiometry distribution (Figure 4).

**5. References**

1.  Berry S and Dean C (2015) Environmental perception and epigenetic memory: Mechanistic insight through FLC. Plant J 83(1):133–148

2.  Wu Z, Fang X, Zhu D, et al (2020) Autonomous pathway: Flowering locus c repression through an antisense-mediated chromatin-silencing mechanism. Plant Physiol 182(1):27–37

3.  Xu C, Wu Z, Duan HC, et al (2021) R-loop resolution promotes co-transcriptional chromatin silencing. Nat Commun 12(1):1790

4.  Macknight R, Bancroft I, Page T, et al (1997) FCA, a gene controlling flowering time in arabidopsis, encodes a protein containing RNA-binding domains. Cell 89(5):737–745

5.  Fang X, Wang L, Ishikawa R, et al (2019) Arabidopsis FLL2 promotes liquid–liquid phase separation of polyadenylation complexes. Nature 569(7755):265–269

6.  McSwiggen DT, Mir M, Darzacq X, et al (2019) Evaluating phase separation in live cells: diagnosis, caveats, and functional consequences. Genes Dev 33(23–24):1619–1634

7.  Sabari BR, Dall'Agnese A, and Young RA (2020) Biomolecular Condensates in the Nucleus. Trends Biochem Sci 45(11):961–977

8.  Emenecker RJ, Holehouse AS, and Strader LC (2020) Emerging Roles for Phase Separation in Plants. Dev Cell 55(1):69–83

9.  Holehouse AS and Pappu R V. (2018) Functional Implications of Intracellular Phase Transitions. Biochemistry 57(17):2415–2423

10. Pliss A, Levchenko SM, Liu L, et al (2019) Cycles of protein condensation and discharge in nuclear organelles studied by fluorescence lifetime imaging. Nat Commun 10(1):455

11. Bayguinov PO, Oakley DM, Shih CC, et al (2018) Modern Laser Scanning Confocal Microscopy. Curr Protoc Cytom 85(1):e39

12. Berthet B and Maizel A (2016) Light sheet microscopy and live imaging of plants. J Microsc 263(2):158–164

13. Lenn T and Leake MC (2012) Experimental approaches for addressing fundamental biological questions in living, functioning cells with single molecule precision. Open Biol 2(6):120090

14. Plank M, Wadhams GH, and Leake MC (2009) Millisecond timescale slimfield imaging and automated quantification of single fluorescent protein molecules for use in probing complex biological processes. Integr Biol 1(10):602–612

15. Chen Y, Müller JD, Ruan Q, et al (2002) Molecular brightness characterization of EGFP in vivo by fluorescence fluctuation spectroscopy. Biophys J 82(1):133–144

16. Wollman AJM, Shashkova S, Hedlund EG, et al (2017) Transcription factor clusters regulate genes in eukaryotic cells. eLife 25(6):e27451

17. Shashkova S, Wollman AJM, Leake MC, et al (2017) The yeast Mig1 transcriptional repressor is dephosphorylated by glucose-dependent and -independent mechanisms. FEMS Microbiol Lett 364(14):fnx133

18. Shashkova S, Nyström T, Leake MC, et al (2021) Correlative single-molecule fluorescence barcoding of gene regulation in Saccharomyces cerevisiae. Methods 193:62–67

19. Wollman AJM, Hedlund EG, Shashkova S, et al (2020) Towards mapping the 3D genome through high speed single-molecule tracking of functional transcription factors in single living cells. Methods 170:82–89

20. Zapata L, Ding J, Willing EM, et al (2016) Chromosome-level assembly of Arabidopsis thaliana Ler reveals the extent of translocation and inversion polymorphisms. Proc Natl Acad Sci U S A 113(28):E4052-60

21. Miller H, Zhou Z, Wollman AJM, et al (2015) Superresolution imaging of single DNA molecules using stochastic photoblinking of minor groove and intercalating dyes. Methods 88:81–88

22. Wollman AJM and Leake MC (2015) Millisecond single-molecule localization microscopy combined with convolution analysis and automated image segmentation to determine protein concentrations in complexly structured, functional cells, one cell at a time. Faraday Discuss 184:401–424

23. Shepherd JW, Higgins EJ, Wollman AJM, et al (2021) PySTACHIO: Python Single-molecule TrAcking stoiCHiometry Intensity and simulatiOn, a flexible, extensible, beginner-friendly and optimized program for analysis of single-molecule microscopy data. Comput Struct Biotechnol J 19:4049–4058

24. Leake MC, Chandler JH, Wadhams GH, et al (2006) Stoichiometry and turnover in single, functioning membrane protein complexes. Nature 443(7109):355–358

25. Rahni R and Birnbaum KD (2019) Week-long imaging of cell divisions in the Arabidopsis root meristem. Plant Methods 15(30)

26. Stoeva D, Göschl C, Corliss B, et al (2017) Long-term confocal imaging of Arabidopsis thaliana roots for simultaneous quantification of root growth and fluorescent signals, In: Plant Genomics: Methods in Molecular Biology, 1610 pp. 169–183

27. Jin X, Lee J-E, Schaefer C, et al (2021) Membraneless organelles formed by liquid-liquid phase separation increase bacterial fitness. Sci Adv (in press)

28. Sun Y, Wollman AJM, Huang F, et al (2019) Single-organelle quantification reveals stoichiometric and structural variability of carboxysomes dependent on the environment. Plant Cell 31(7):1648–1664

29. Cosgrove J, Novkovic M, Albrecht S, et al (2020) B cell zone reticular cell microenvironments shape CXCL13 gradient formation. Nat Commun 11(1):3677

30. Miller H, Cosgrove J, Wollman AJM, et al (2018) High-speed single-molecule tracking of CXCL13 in the B-follicle. Front Immunol 9:1073

31. Stracy M, Wollman AJM, Kaja E, et al (2019) Single-molecule imaging of DNA gyrase activity in living Escherichia coli. Nucleic Acids Res 47(1):210–220

32. Badrinarayanan A, Reyes-Lamothe R, Uphoff S, et al (2012) In vivo architecture and action of bacterial structural maintenance of chromosome proteins. Science 338(6106):528–531

33. Dresser L, Hunter P, Yendybayeva F, et al (2021) Amyloid-β oligomerization monitored by single-molecule stepwise photobleaching. Methods 193:80–95

34. Syeda AH, Wollman AJM, Hargreaves AL, et al (2019) Single-molecule live cell imaging of Rep reveals the dynamic interplay between an accessory replicative helicase and the replisome. Nucleic Acids Res 47(12):6287–6298

35. Reyes-Lamothe R, Sherratt DJ, and Leake MC (2010) Stoichiometry and architecture of active

DNA replication machinery in escherichia coli. Science 328(5977):498–501

36. Meschichi A and Rosa S (2021) Visualizing and Measuring Single Locus Dynamics in Arabidopsis thaliana, In: Arabidopsis Protocols: Methods in Molecular Biology, 2200 pp. 213–224

37. Wu X and Hammer JA (2021) ZEISS airyscan: optimizing usage for fast, gentle, super-resolution imaging, In: Confocal Microscopy: Methods in Molecular Biology, 2304 pp. 111–130

**Acknowledgements**

Thanks to Geng-Jen Jang and Caroline Dean (John Innes Centre, Norwich) for donating *Arabidopsis* seeds for the FCA-EGFP fusion construct. This work was supported by the EPSRC (EP/T002166/1).